# Report on the Conference on Ethical and Responsible Design in the National AI Institutes: A Summary of Challenges

Authors: Sherri Conklin, Sue Bae (AI4OPT), Gaurav Sett, Michael Hoffmann (AI-ALOE), and Justin Biddle (AI4OPT)[*]

Contributors: Carrie Alexander (AIFS), Ann Bostrom (AI2ES), Margaret Burnett (AgAID), Jeff Bush (ISAT), Michael Alan Chang (ISAT), Nathan Colaner (AI Institute in Dynamic Systems), Phillip L Davis (AI2ES), Imme Ebert-Uphoff (AI2ES), Ashok Goel (AI-ALOE), Venu Govindaraju (AI Institute for Exceptional Education), Ashish Hingle (AIIRA), Aditya Johri (AIIRA), Sadia Khan (ICICLE), Christina Kumler (AI2ES), Collin F. Lynch (ENGAGE AI), Brian Magerko (AI-ALOE), Jessie Micallef (IAIFI), Ole Molvig (ENGAGE AI), Srinivasan Parthasarathy (AI-EDGE), Russell Perkins (AI-CARING), Beth Plale (ICICLE), Paul Robinette (AI-CARING), Neelima Savardekar (ICICLE), Sanaz Ahmadzadeh Siyahrood (AI-ALOE), Jason Stock (AI2ES), Miranda White (AI2ES), Jinjun Xiong (AI Institute for Exceptional Education)

## Introduction

The development and proliferation of artificial intelligence (AI) systems has generated both optimism and concern. AI has the potential to benefit humanity in significant ways, but it also brings risks – including risks of discrimination, inequality, labor disruptions, degradation of human autonomy, political instability, and many others. The U.S. National Artificial Intelligence Research and Development Strategic Plan, updated in 2023, identifies nine key strategies for national AI research and development (R&D), one of which is to "understand and address the ethical, legal, and societal implications of AI." The establishment of the U.S. National AI Institutes represents a commitment to advance AI R&D that delivers on these strategies. As such, it is important that the National AI Institutes are structured to promote the ethical and responsible design of AI systems. Many of the National AI Institutes are attempting to promote ethical and responsible design, but they face significant challenges to do this effectively.

In May 2023, the Georgia Tech Ethics, Technology, and Human Interaction Center (ETHIC[x]) organized the Conference on Ethical and Responsible Design in the National AI Institutes.

---

[*] *Corresponding author*: Justin B. Biddle, School of Public Policy, Georgia Institute of Technology, Atlanta, Georgia 30332-0345. *Email*: justin.biddle@pubpolicy.gatech.edu

Representatives from all of the National AI Research Institutes that had been established as of January 2023 were invited to attend; researchers representing 14 Institutes attended and participated. The conference focused on three questions: What are the main challenges that the National AI Institutes are facing with regard to the responsible design of AI systems? What are promising lines of inquiry to address these challenges? What are possible points of collaboration? In the course of the conference, a revised version of the first question became a focal point: What are the challenges that the Institutes face in identifying ethical and responsible design practices and in implementing them in the AI development process? *This document summarizes the challenges that representatives from the Institutes in attendance highlighted.*

These challenges fall into four inter-related categories: (1) challenges associated with the organization of the National AI Institutes – how they should be organized so as to facilitate responsible and ethical design; (2) challenges associated with the adoption, implementation, and evaluation of frameworks for ethical and responsible design; (3) challenges associated with how National AI Institutes engage with society, and (4) challenges associated with how National AI Institutes coordinate with each other. All of the challenges discussed in this document were directly contributed by attendees, and observations about these challenges were discussed throughout the conference. This document is correspondingly a collaborative project with contributions from every participating Institute.

To write this report, the authors documented discussions that took place at the conference and drafted a summary report, which was distributed to the conference participants – including those who are listed as contributors – for feedback. The authors then incorporated this feedback into the final report.

**1. Organizing the National AI Institutes for Ethical and Responsible Design**

One major challenge experienced across the Institutes pertained to the structures and cultures of the Institutes themselves. Notwithstanding the fact that some Institutes have dedicated ethics thrusts, questions remain regarding how to effectively integrate ethics work into the Institutes. Three challenges appeared to garner the most attention during the conference. These are: (1) ethics siloing; (2) incentivizing ethics work; and (3) distributing ethical responsibility.

*1.1 The Silo Problem*

Ethics within many Institutes is siloed to a significant extent. In many cases, ethics work is conducted by an individual or small team, isolated to one project, and not, more generally, integrated into work across the Institute. Some noted, for example, how limited contact between researchers with an ethics focus and the rest of the Institute made it difficult to communicate ethics-related issues to the other teams or individuals within the Institute. At some Institutes, ethics is confined to one research thrust. In several such cases, ethics work focuses on aspects of the Institute's public-facing research that are anticipated to impact users and stakeholders. Ethics in these cases often focuses on education or communication. In some Institutes, ethics is siloed to education or outreach teams and away from AI development teams. Some reported that it is especially difficult to convince developers that ethics is relevant not only to how AI systems are used but also to how they are designed; even if developers are well intentioned, ethical values can get built into AI systems and impact society in ways that are not intended or foreseen.

In some cases of ethics siloing, considerations of ethics responsibility was reported to be an afterthought to the Institute's AI research and/or development goals. This might, for example, take the form of introducing an ethics thrust or team after the Institute had already been planned, which reflects consideration of ethics only late in development and planning of the Institute.

*1.2 Incentivizing Ethics Work*

Many conference attendees expressed a need for improved incentive structures for faciliating engagement and ethical reflection and practice. For example, several participants noted that ethics is not always considered a part of the Institute's research agenda. Instead, ethics work is in some cases classified as service rather than research – especially when dealing with concerns about diversity, equity, and inclusion (DEI), which may mean that researchers receive reduced credit for their work or that their work is seen as less important than more technical work. Some noted that this challenge may be systemic to the broader academic environment in which Institutes operate. It was also noted that implementing ethical design practices may introduce a burden of additional work for AI system designers, which, while benefiting other stakeholders, might not have a clear benefit for the designers themselves. While we believe that designers have moral obligations to consider the societal consequences of their work, the concern about additional burdens is relevant to the issue of incentive structures and might play a role in undermining ethics engagement.

*1.3 Distributing Ethical Responsibility*

Many participants reported challenges concerning how to distribute responsibility for ethical decision making within an Institute and how to operationalize the scope of the Institute's ethics mission. For example, some wondered whether ethical aims within an Institute should be realized by recruiting an expert ethicist (or team of ethicists) or by promoting a culture of ethics that pervades the Institute, such that every member of an Institute is encouraged to think systematically about ethics. Each strategy for distributing ethical responsibility presents its own issues. The former, it would seem, could contribute to the sort of ethics siloing discussed above. The latter, one might worry, asks team members who may not have extensive experience with ethics-related research – and who may not recognize the ethical relevance or importance of technical decisions they are making – to nonetheless make judgments that they are not qualified to make.

## 2. Ethics Frameworks

A second major challenge reported by many participants involved ethics frameworks. We use the term "ethics framework" to include guidelines, principles, or standards for ethical and responsible design of AI systems. Ethics frameworks, in this sense, go beyond institutionally mandated ethics requirements such as Responsible Conduct of Research (RCR) training and Institutional Review Board (IRB) oversight of human subjects research. Institutes reported issues in four areas: (1)

selecting ethical frameworks; (2) implementing ethical frameworks; (3) assessing ethical frameworks once they are implemented; and (4) acting at the cutting edge of ethical research or at the boundaries of a framework's scope.

## 2.1. Selecting a Framework

Ethics frameworks are potentially effective ways of implementing ethical and responsible design practices, mitigating risks that AI systems could pose, and promoting societal benefits of AI. Some of the Institutes represented at the conference had already selected and implemented an ethics framework. Examples of frameworks that have been adopted by some National AI Institutes are the *IEEE Recommended Practice for Assessing the Impact of Autonomous and Intelligent Systems on Human Well-Being* (IEEE 7010-2020; used by AI-ALOE) and GenderMag and InclusiveMag (www.gendermag.org, used by AgAID).[1] The National Institute of Standards and Technology (NIST) AI Risk Management Framework (AI RMF) was also highlighted as a potentially useful framework. Some National AI Institutes were developing ethics frameworks that are tailored to their specific Institutes (e.g., AIFS). Other Institutes had not yet made decisions about whether to select an ethics framework and, if so, how to do this. Despite the variation between Institutes, most reported challenges associated with the selection of an ethics framework. Some of these challenges result from differences between ethics frameworks (e.g., between general, off-the-shelf frameworks versus specifically-tailored frameworks) and differences between the needs and goals of various Institutes and projects.

An assessment of an Institute's needs and goals is sometimes a prerequisite for selecting an ethics framework. For example, Institutes that allow for a great deal of independent research may tolerate more flexible frameworks whereas Institutes with more constraints (e.g., within universities or by external partners) may require more structured frameworks. Also, the resources available to an Institute, in terms of funding and personnel, will likely impact the type of framework selected. Institutes with dedicated ethics researchers may be equipped to incorporate a more comprehensive ethics framework, yet the efforts of these dedicated ethics personnel could be quickly undermined if insufficient funds or other key resources are allocated to this work. In addition, as we have already noted, the impact of the ethics work could be diminished if there is siloing within an Institute, so consideration of the Institute's overall structure is another factor to consider when selecting the framework.

One question that several representatives reported grappling with is whether to select an off-the-shelf framework or to create and tailor a framework for their particular Institute. Examples of ready-made frameworks include the NIST AI RMF and IEEE 7010-2020, each of which articulates ethical design practices and risk management strategies for AI in general. When a pre-existing framework covers an Institute's research area, implementation of that framework could be relatively easy, because the framework creators will likely have considered many of the important ethical dimensions of that work. However, some Institutes will have specialized needs and goals, which might justfify the creation of a new and tailored framework.

---

[1] Mendez, C., L. Letaw, M. Burnett, S. Stumpf, A. Sarma and C. Hilderbrand (2019). "From GenderMag to InclusiveMag: An Inclusive Design Meta-Method," *2019 IEEE Symposium on Visual Languages and Human-Centric Computing (VL/HCC)*, Memphis, TN, USA: 97-106. DOI: 10.1109/VLHCC.2019.8818889. For the work of AI-ALOE see https://aialoe.org/ai-ethics/.

Cutting across this concern is the problem of ethics expertise. Many Institute representatives expressed concern about anticipating emergent issues at an early stage in the development of an AI system. This can be compounded when AI researchers lack experience with user engagement, co-production of research, or co-design with stakes- and rights-holders, and with the broader social issues that bear on their work. Without this knowledge of ethically salient considerations for the communities potentially served or affected by an AI system, it is challenging to anticipate all of the ethics-related needs that will emerge. It can be difficult for researchers to determine whether some framework will fit the needs of an Institute or specific project within the Institute without interdisciplinary ethics experience, experience with the framework, or useful proxies for experience, such as detailed case-studies. Developing a new and tailored framework, which offers the benefit of customizeability and project-specific ethics sensitivity, requires additional domain-specific background knowledge and substantial resources. As a result, the amount of ethics expertise or expertise with ethics frameworks within the Institute will impact which options are reasonably available.

## 2.2. Implementing a Framework

After selecting an ethical framework, researchers face further challenges in implementing it responsibly. A significant challenge reported by conference attendees was how to translate a framework – which is comprised of general principles or guidelines – into concrete practices. For example, some ethical frameworks require participatory engagement – including identification of relevant stakeholders and engagement with them – but do not specify the precise methods or scope of effort required for facilitating this engagement (e.g., focus groups, surveys, structured interviews, or the like). Decisions about how to operationalize a requirement such as participatory engagement are, in such cases, left to the discretion of the researchers or Institutes implementing them. Additionally, some ethical frameworks require that researchers identify and measure potential risks and benefits of AI systems. However, they might not specify precisely which risks and benefits to consider, which methods are most well-suited for measuring them (e.g., quantitative or qualitative), or whether doing so is desirable or even possible. Again, such decisions are left to the discretion of individual researchers or Institutes.

By highlighting the challenges associated with implementing an ethics framework, we are not suggesting that researcher discretion can, or should, be eliminated. Researching and developing AI systems requires judgment on the part of investigators, and it is neither feasible nor desirable to attempt to override the ability of researchers or Institutes to make such discretionary judgments. But how best to support researchers and Institutes – whether by training individual researchers to make judgments in a more ethically-informed manner, or by creating processes that can better assist them in this endeavor, or by pursuing some other strategy – is an open and challenging question.

In addition, some attendees expressed concerns about how onerous the process of implementing a framework can be, which can discourage initiating the process in the first place. Some were concerned about the ways in which new responsibilities translate to new burdens for the researchers within the Institute and potentially for partners on whose collaboration they depend.

## 2.3. Evaluating Frameworks

After implementing an ethics framework, evaluating its impact and efficacy poses additional challenges. Rigorously determining a framework's influence on the design of an AI technology, as well as its fit within an Institute, requires robust measurement. But it is an open research question – both within National AI Insitutes and in the broader research community – which metrics are most helpful in assessing framework effectiveness.

Success metrics should measure how well the framework achieves an Institute's ethics needs and goals, especially with regard to the ethical impact of the AI technology itself. For example, an Institute may need to consider how well a framework supports the Institute in accurately predicting and obviating harmful outcomes, such as significantly impinging on the interests or well-being of stakeholders. An Institute may also need to evaluate how well the framework adapts to unexpected events and outcomes. If an Institute foresees that an AI technology is likely to significantly impinge on the interest of stakeholders, it might consider the role of the framework in identifying those risks and in facilitating the redesign of the technology, so that the risks of these harm is significantly reduced. When Institutes fail to foresee these outcomes, they might consider what guidance the framework offers on managing such failures, as even diligent framework application may result in unforeseen issues, and on communicating risks and failures transparently.

## 2.4. Acting at the Edge of a Framework

One issue that cuts across the problems of selecting, implementing, and evaluating an ethics framework involves acting at the edges of a framework's scope. This issue arises because many Institutes are dealing with unresolved ethical problems for which there are no agreed-upon solutions. In other cases, Institutes might be advancing the boundaries of their respective fields and encountering novel ethics problems. While navigating ethical gray areas that remain ongoing areas of research in the field of applied and practical ethics, or that emerge where ethics guidelines and interests misalign, the needs of these Institutes might surpass the capabilities of existing frameworks. Representatives of some Institutes expressed concerns about their responsibilities when it comes to acting and making-decisions at the edges of available expertise. Some others have worked to address this issue in their own Institutes.[2]

This challenge makes it difficult to anticipate all of the important ethical questions that an Institute will face while developing an AI tool and contributes to the challenge of selecting an ethics framework comprehensive and flexible enough to cover the full scope of the Institute's ethics mission. Regardless of the problem's source, some Institute representatives were concerned about the extent to which it is their responsibility to select frameworks that can identify solutions to open research questions in ethics, especially when making progress on these issues might impede on the Institute's ability to achieve its primary purpose. This challenge also makes it difficult for an Institute to determine the scope of its ethics mission. In particular, researchers at several Institutes questioned whether Institutes should be held responsible for solving ethics-related issues that dedicated researchers in practical ethics are still grappling with. At the same time, some researchers within the AI Institutes also reported feeling disempowered by the limits of what they can reasonably accomplish or address.

---

[2] See, for example, Alexander, C. S., M. Yarborough, & A. Smith (2023), "Who is Responsible for 'Responsible AI'?: Navigating Challenges to Build Trust in AI Agriculture and Food System Technology," *Precision Agriculture* 25 (6097): 1-40.

**3. AI Institutes and Society**

As already noted, the AI Institutes must navigate a complex environment with diverse stakeholders. Engagement with various stakeholders can shape how AI research is conducted, but social and economic pressures may not always favor ethical and responsible design practices. This section explores the ethical challenges that social and economic pressures place on Institutes, in addition to some other important issues relating to social engagement. These ethical challenges involve concerns about: (1) counterproductive social and economic pressures; and (2) stakeholder engagement.

*3.1. Counterproductive Social and Economic Pressures*

Different stakeholders have different interests and values. Stakeholders include users, investors, industry collaborators, publics, politicians, and researchers, but who counts as a stakeholder varies widely among the Institutes. For the Institutes creating AI tools that interface with vulnerable populations, such as children, the elderly, or migrant workers in the agricultural industry, engagement with these stakeholders can have serious and life-altering ethical implications. These Institutes should strive to utilize forms of inquiry and to implement design processes that comprehensively consider the legitimate perspectives of diverse stakeholders. However, not all stakeholder engagement is productive for ethical decision-making during the design process, and some social and economic pressures are at cross-purposes with achieving ethical goals.

One group of stakeholders who may have a disproportionate impact on project outcomes is industry partners. Institutes frequently collaborate with industry, and Institute representatives expressed concern that some industry partners might deprioritize ethical review in an effort to be first-to-market. Responsible development may also demand resources and actions that could cut into profits. Institutes may lose control over technologies after deployment, and some participants raised questions about ensuring responsible use post-deployment, which may require continued engagement and accountability procedures. While this latter concern applies to Institute collaborations more generally, it was raised most pointedly in the context of industry.

Another group of stakeholders that was subject of discussion is researchers within the Institutes themselves. While some Institutes pursue research that is directly relevant to social and ethical concerns – such as Institutes that focus on the development of AI tools for education – other Institutes pursue upstream research with social impacts that are less readily apparent. It might be more challenging for Institutes in the latter category to locate AI ethics concerns than in the former.

If ethics is viewed through a lens that focuses on direct social impact, it can be difficult to characterize and address AI ethics concerns such as downstream or secondary applications of an AI tool, including how the tool will be used by research partners like the military or industry. Researchers face challenging questions involving assessing potential misuse, including of dual-use technologies, and balancing openness versus caution in publication.

This lens can also contribute to the attitude that ethics is not relevant to "pure" or "basic" scientific research – that it is relevant only to the social sciences or to applied and socially engaged projects – which can contribute to the aforementioned siloing effects and discourage new researchers from seeing ethics as a priority. Several Institute representatives called for greater awareness of the relevance and importance of ethics in research and that such awareness be reflected in promotion and tenure policies as well as in the priorities of funding agencies.

### 3.2. Stakeholder Engagement

Institutes have obligations to engage with external stakeholders and communicate with them about aspects of their AI research. To promote stakeholder trust, Institutes will typically need to elicit feedback from stakeholders in the AI development process and communicate how this feedback was incorporated into the final product prior to adoption or deployment. Similarly, when communicating findings, transparency about data bias, technology effectiveness, or societal risks might be necessary.

Effective, two-way communication between Institutes and stakeholders requires proper intent, motivation and resources on all sides. In some situations, stakeholders might need to be educated about aspects of an AI technology about which they are misinformed; at the same time, AI researchers and Institutes should not assume that they understand the interests, perspectives, and needs of the communities that they affect, without spending the time and resources required to do this. Improving communication between Institutes and stakeholders might require a variety of efforts, from increasing the interpretability and explainability of AI systems to experimenting with new tools and processes for community engagement and co-production.

### 4. Coordination between AI Institutes

Finally, workshop attendees repeatedly remarked on how a lack of coordination between AI Institutes poses a challenge for the ethical and responsible design of AI. Prior to this conference, there had been relatively little coordination about ethical and responsible design between the National AI Institutes. As a result, many Institutes that are grappling with similar ethics issues manage these challenges on their own. The previously-discussed siloing is inefficient; information about successful or failing strategies for managing ethics concerns is not shared, which results in duplicated efforts and lost time. This conference was a first important step to improve coordination and collaboration among the National AI Institutes to increase the efficacy and efficiency of our work. As proposed at the conference, we have founded a Special Interest Group (SIG) on Ethics and Trustworthiness, as a part of the AI Institutes Virtual Organization (AIVO) to help facilitate coordition about ethics and trustworthiness between the National AI Institutes. This SIG includes representatives of National AI Institutes that had not yet been established at the time our conference was planned, including the NSF Institute for Trustworthy AI in Law and Society (TRAILS).

Institutes also expressed a desire for a collective library of case studies, frameworks, tools and best practices that could inform approaches across Institutes. Codifying successful solutions and ethical deliberations into an organized knowledge base could amplify their impact. Institutes could also share experiences through ongoing exchanges. Contact points for various expertise areas could connect researchers to advise on issues faced. A panel of experts across Institutes could provide guidance on novel problems. Workshops to teach successful techniques could propagate solutions.

**Acknowledgements**